\documentclass[%
 reprint,
superscriptaddress,
 amsmath,amssymb,
 aps,
showkeys,
]{revtex4-2}

\usepackage{graphicx}
\usepackage{dcolumn}
\usepackage{bm}
\usepackage[italicdiff]{physics}
\usepackage{amsmath, mathtools, amssymb, ascmac, fancybox, xcolor, amsfonts,mathrsfs}
\usepackage[colorlinks=true,linkcolor=teal,citecolor=teal,urlcolor=blue]{hyperref}
\usepackage{breakurl}

\begin{document}

\preprint{APS/123-QED}

\title{Conserved quantities in a bosonic tight-binding chain with non-Hermitian quartic terms}


\author{Tetsuya Iwasaki}
\affiliation{Department of Physics, Graduate School of Science, The University of Tokyo, Tokyo~113-0033, Japan}

\author{Hosho Katsura}
\affiliation{Department of Physics, Graduate School of Science, The University of Tokyo, Tokyo~113-0033, Japan}
\affiliation{Institute for Physics of Intelligence, The University of Tokyo, Tokyo 113-0033, Japan}
\affiliation{Trans-Scale Quantum Science Institute, The University of Tokyo, Tokyo 113-0033, Japan}

\date{\today}

\begin{abstract}
We investigate local conserved quantities in non-Hermitian bosonic lattice systems whose underlying Yang--Baxter structure remains unclear.
Focusing on a one-dimensional bosonic chain with quartic interactions of creation operators,
we provide a new representation of the local conserved quantities previously constructed by Sanatani and Shiraishi.
Using Fourier transformation and trigonometric identities, we
systematically derive these conserved quantities
and show directly that they form a mutually commuting family.
We further extend the construction to interactions extending beyond a single site, cubic interactions, asymmetric hopping, and an on-site potential.
Our results provide a unified framework for constructing and characterizing local conserved quantities in this class of non-Hermitian many-body systems.
\end{abstract}

\maketitle


\section{\label{sec:intro}Introduction}
Quantum integrable systems provide a special class of many-body systems that exhibit remarkable mathematical structures and exact solutions \cite{KorepinBogoliubovIzergin,Takahashi,Baxter,Sutherland,Gaudin}.
A defining feature of quantum integrability is the existence of an extensive set of conserved quantities,
which strongly constrains the dynamics and distinguishes integrable systems from generic many-body systems.
The presence of these conserved quantities has profound consequences for the thermalization \cite{Rigoletal,CassidyRigolClark,Pozsgay}, transport \cite{ZotosNaefPrelovsek,Zotos,FujimotoKawakami,Bertinietal}, and hydrodynamic properties \cite{Castro-AlvaredoDoyonYoshimura,BertiniColluraDeNardisFagotti,Pozsgay2} of integrable systems.
The construction and characterization of these conserved quantities have therefore been an important subject of study in quantum integrability.
Two main approaches have been developed for constructing conserved quantities.
One is a systematic construction based on the Yang--Baxter equation,
in which conserved quantities are generated from the associated transfer matrix \cite{DolanGrady,KorepinBogoliubovIzergin,Baxter}.
Alternatively, conserved quantities can be constructed directly by solving the conservation condition for local operators \cite{NozawaFukai,YamadaFukai,Fukai1,Chiba,Fukai2,FukaiYamada}.
Such direct constructions have also been used to establish nonintegrability by demonstrating the absence of the required conserved quantities \cite{Shiraishi1,Chiba,ChibaYoneta,ParkLee,Shiraishi2,SanataniChibaShiraishi,Hokkyo2,HokkyoYamaguchiChiba,ParkLee2,Chiba2,ShiraishiTasaki,Shiraishi3,FutamiTasaki,ShiraishiYamaguchi,FanHaoChenZhangWangKorepin,Futami,YamaguchiIshii,SengokuWatanabe}.
The hierarchy of conserved quantities can also be understood through the boost operator,
which connects conserved quantities of successive orders \cite{SogoWadati,Hokkyo}.
In particular, the existence of a conserved quantity of a low-order can lead to the automatic construction of higher-order conserved quantities \cite{GrabowskiMathieu,Hokkyo,ShiraishiYamaguchi},
giving rise to an all-or-nothing structure.

While integrable systems have traditionally been studied primarily in Hermitian settings, integrability has also been established in non-Hermitian quantum systems \cite{baxter1989simple,pasquier1990common,AlcarazDrozHenkelRittenberg,AlbertiniDahmenWehefritz,BogoliubovNassar,FukuiKawakami,Bogoliubov,NakagawaKawakamiUeda2,BucaBookerMedenjakJaksch,NakagawaKawakamiUeda,YamamotoNakagawaTezukaUedaKawakami,MaoHaoPan,KattelPasnooriAndrei,IshiguroSatoNishinari,ZhengQiaoWangCaoChen,KattelPasnooriPixleyAndrei,MaityPadmanabhanKorepin}.
A notable example is the Bose--Hubbard model with unidirectional hopping,
whose integrability can be understood within the Yang--Baxter framework \cite{ZhengQiaoWangCaoChen}.
This provides an example of a non-Hermitian bosonic system retaining the conventional algebraic structure of quantum integrability.
More recently, Sanatani and Shiraishi investigated a broad class of one-dimensional bosonic lattice models with symmetric hopping and on-site interactions
and identified several models admitting nontrivial local conserved quantities in the non-Hermitian regime \cite{SanataniShiraishi}.
Their analysis includes models exhibiting different structures in their conserved quantities,
including cases where the usual all-or-nothing structure is not observed.
In contrast to the unidirectional Bose--Hubbard model, the underlying Yang--Baxter structure of these models remains unclear.
Consequently, the underlying structure responsible for these conserved quantities is not yet fully understood.
It is also unclear whether the conserved quantities found in these models can be organized into a mutually commuting family.
Understanding the underlying structure of these conserved quantities is therefore an important open problem.

In this work, we focus on a specific model from the broad class of models investigated by Sanatani and Shiraishi and revisit its conserved quantities using Fourier transformation and trigonometric identities.
This approach allows us to reproduce the family of local conserved quantities found in Ref.~\cite{SanataniShiraishi}.
Moreover, their mutual commutativity becomes manifest from the construction.
We further show that the same framework can be extended to models with interactions beyond on-site terms, fermionic systems, and systems with asymmetric hopping.
While the underlying Yang--Baxter structure, if any, remains to be clarified,
our results provide a unified framework for constructing and understanding these conserved quantities.

The remainder of this paper is organized as follows.
In Sec.~\ref{sec:construction}, we introduce the bosonic models with quartic interactions and construct its local conserved quantities using Fourier transformation and trigonometric identities.
In Sec.~\ref{sec:generalize}, we extend the construction to a wider class of models, including interactions extending beyond a single site, cubic interactions, asymmetric hopping, and an on-site potential.
Finally, we summarize our results and present an outlook for future work in Sec.~\ref{sec:conclusion}.
Technical details on the trigonometric identities and the construction of the conserved quantities are provided in the Appendices.

\section{\label{sec:construction}model and main results}

\subsection{\label{subsec:setting}Model and Fourier transformation}
We consider a translationally invariant bosonic chain with $L$ sites.
Throughout this work, we restrict ourselves to even $L$.
The Hamiltonian is given by
\begin{equation}
       H = T + V,
       \label{eq:Ham1}
\end{equation}
with
\begin{equation}
        T = \sum_{j=1}^{L}\qty(b^\dagger_jb_{j+1} + b^\dagger_{j+1}b_{j}), \quad 
        V= c\sum_{j=1}^{L}\qty(b_j^\dagger)^4,\label{eq:ham4}
\end{equation}
where $b^\dagger_j$ ($b_j$) is the boson creation (annihilation) operator at site $j$, $c$ is a constant, and periodic boundary conditions are imposed.
This Hamiltonian is a special case of integrable models studied in Ref.~\cite{SanataniShiraishi}.

We then perform a Fourier transformation,
\begin{equation}
    b_j = \frac{1}{\sqrt{L}}\sum_{p}e^{ipj}b_p, \quad b_j^\dagger = \frac{1}{\sqrt{L}}\sum_{p}e^{-ipj}b_p^\dagger,
\end{equation}
where $p = 2\pi m/L$ with $m = 0,1,2,\ldots, L-1$, defined modulo $2\pi$.
The Hamiltonian \eqref{eq:ham4} is expressed in Fourier space as
\begin{align}
    &T = \sum_{p}2\cos(p)n_p,\\
    &V = \frac{c}{L}\sum_{p_1,p_2,p_3,p_4}\delta_{p_1+p_2+p_3+p_4,0}\,b^\dagger_{p_1}b^\dagger_{p_2}b^\dagger_{p_3}b^\dagger_{p_4}.
\end{align}
Here, $n_p = b^\dagger_p b_p$ is the number operator of mode $p$, and $\delta_{p,q}$ denotes the Kronecker delta modulo $2\pi$.
An important observation is that the terms within each of the two sums mutually commute:
\begin{equation}
    [n_p,n_q] = 0,\quad [b^\dagger_{p_1}b^\dagger_{p_2}b^\dagger_{p_3}b^\dagger_{p_4},b^\dagger_{q_1}b^\dagger_{q_2}b^\dagger_{q_3}b^\dagger_{q_4}] = 0.\label{eq:com-of-each-element}
\end{equation}
In the following, we construct a family of conserved quantities of this model explicitly.

\subsection{\label{subsec:construction}Conserved quantities}

We first present the family of mutually commuting conserved quantities, which includes the Hamiltonian $H$ in Eq. \eqref{eq:Ham1} as a particular member:
\begin{equation}
    Q^{(n)} = T^{(n)} + V^{(n)},\quad 1\leq n\leq L/2,
    \label{eq:Qn4}
\end{equation}
where 
\begin{align}
    &T^{(n)} = \sum_{p}\epsilon^{(n)}(p)n_p,\\
    &V^{(n)} = \frac{c}{L}\sum_{p_1,p_2,p_3,p_4}g^{(n)}(p_1,p_2,p_3,p_4)\notag\\
    &\quad \times\delta_{p_1+p_2+p_3+p_4,0}\,b^\dagger_{p_1}b^\dagger_{p_2}b^\dagger_{p_3}b^\dagger_{p_4}.
\end{align}
\footnote{The range of $n$ is restricted by the requirement of locality.
For the models considered here,
the conserved quantities satisfy $Q^{(n)} = Q^{(n+L)}$.
Moreover, for symmetric hopping, $Q^{(L-n)}$ is not independent of $Q^{(n)}$, up to lower-order conserved quantities.
These relations follow from the construction of the conserved quantities given below.}
Here, $\epsilon^{(n)}(p)$ and $g^{(n)}(p_1,p_2,p_3,p_4)$ are given by
\begin{align}
    &\epsilon^{(n)}(p) = \begin{cases}
        2\cos(np) & (n\ \text{odd})\\
        2i\sin(np) & (n\ \text{even})
    \end{cases},\label{eq:en4}\\
    &g^{(n)}(p_1,p_2,p_3,p_4) = \frac{\sum_{l=1}^{4}\epsilon^{(n)}(p_l)}{\sum_{l=1}^{4}2\cos(p_l)}.\label{eq:gn4}
\end{align}
Note that $Q^{(1)} = H$.
The ratio 
defining $g^{(n)}$ has an apparent singularity when the denominator vanishes.
However, the numerator vanishes whenever the denominator does, so that the apparent singularity is removable.
This cancellation will be crucial for establishing the locality of $Q^{(n)}$ below.

We now show that the operators $Q^{(n)}$, $n=1,2,\ldots,L/2$, form a mutually commuting family. 
Using Eq.~\eqref{eq:com-of-each-element}, we obtain
\begin{equation}
    [Q^{(n)},Q^{(m)}] = [T^{(n)},V^{(m)}] - [T^{(m)},V^{(n)}].\label{eq:com-QQ}
\end{equation}
The commutator $[T^{(n)},V^{(m)}]$ is given by
\begin{align}
    &[T^{(n)},V^{(m)}] \\
&= \frac{c}{L}\sum_{p_1,p_2,p_3,p_4}\frac{\qty[\sum_{l=1}^{4}\epsilon^{(n)}(p_l)]\qty[\sum_{l=1}^{4}\epsilon^{(m)}(p_l)]}{\sum_{l=1}^42\cos(p_l)}\notag\\
    &\quad\times
\delta_{p_1+p_2+p_3+p_4,0}~b^\dagger_{p_1}b^\dagger_{p_2}b^\dagger_{p_3}b^\dagger_{p_4},
\end{align}
which is symmetric under the exchange $n\leftrightarrow m$.
It therefore follows from Eq.~\eqref{eq:com-QQ} that
\begin{equation}
    [Q^{(n)},Q^{(m)}] = 0.
\end{equation}
In particular, $Q^{(n)}$ are conserved quantities of $H$.

We next show that $Q^{(n)}$ is local.
Although $g^{(n)}$ is defined as a ratio, it is in fact a trigonometric polynomial.
Under the momentum-conservation condition $p_1+p_2+p_3+p_4 = 0$ (mod $2\pi$),
the numerator contains the denominator as a factor:
\begin{equation}
    \sum_{l=1}^{4}\epsilon^{(n)}(p_l) = \qty(\sum_{l=1}^42\cos p_l)f^{(n)}(p_1,p_2,p_3,p_4),\label{eq:trig-form-1}
\end{equation}
where $f^{(n)}$ is a trigonometric polynomial whose degree is bounded by $n-1$, independently of $L$. 
The proof of formula \eqref{eq:trig-form-1} is given in Appendix~\ref{app:quartic}.
We note in passing that a related approach to constructing conserved quantities and establishing their locality through trigonometric identities was employed in the construction of quasi-charges for perturbed free-fermion systems \cite{PozsgaySharipovTiutiakinaVona}.

For illustration, we explicitly give $f^{(n)}$ for the first few values of $n$:
\begin{align}
    f^{(2)} &= \sum_{l=1}^{4}2i\sin(p_l),\\
    f^{(3)} &= \sum_{l=1}^42\cos(2p_l)+\sum_{1\leq l<m\leq 4}e^{i(p_l+p_m)} \notag\\&-\sum_{l,m=1}^4e^{i(p_l-p_m)} + 3,\\
    f^{(4)} &= \sum_{l=1}^42i\sin(p_l)\sum_{m=1}^42\cos(2p_m).
\end{align}
Substituting these expressions into Eq.~\eqref{eq:Qn4} and transforming back to real space, we obtain
\begin{align}
    Q^{(2)} &= \sum_{j=1}^L\qty[b^\dagger_jb_{j+2}-b^\dagger_{j+2}b_j]\notag\\&+c\sum_{j=1}^L\qty[4b^\dagger_{j}\qty(b^\dagger_{j+1})^3-4\qty(b^\dagger_j)^3b^\dagger_{j+1}],
\end{align}
\begin{align}
    Q^{(3)}&= \sum_{j=1}^L\qty[b^\dagger_jb_{j+3} + b^\dagger_{j+3}b_j]\notag\\
    &+c\sum_{j=1}^{L}\left[4b^\dagger_j\qty(b^\dagger_{j+2})^3+4\qty(b^\dagger_{j})^3b^\dagger_{j+2} + 6\qty(b^\dagger_{j})^2\qty(b^\dagger_{j+1})^2 \right. \notag\\
    &-\left.12\qty(b^\dagger_{j})\qty(b^\dagger_{j+1})^2\qty(b^\dagger_{j+2}) - \qty(b^\dagger_j)^4\right],
\end{align}
\begin{align}
    &Q^{(4)} = \sum_{j=1}^L\qty[b^\dagger_{j}b_{j+4}-b^\dagger_{j+4}b_j]\notag\\
    &\quad + c\sum_{j=1}^L\left[4b^\dagger_j\qty(b^\dagger_{j+3})^3-4\qty(b^\dagger_j)^3b^\dagger_{j+3}-4b^\dagger_j\qty(b^\dagger_{j+1})^3\right.\notag\\
    &+4\qty(b^\dagger_j)^3b^\dagger_{j+1} + 12b^\dagger_jb^\dagger_{j+1}\qty(b^\dagger_{j+2})^2 - 12\qty(b^\dagger_{j})^2b^\dagger_{j+1}b^\dagger_{j+2}\notag\\
    &\quad\left.-12b^\dagger_{j}\qty(b^\dagger_{j+2})^2b^\dagger_{j+3} + 12b^\dagger_{j}\qty(b^\dagger_{j+1})^2b^\dagger_{j+3}\right].
\end{align}
Note that $Q^{(2)}$, $Q^{(3)}$, and $Q^{(4)}$ are equal to $Q_3$, $Q_4-Q_2$, and $Q_5-Q_3$ in Ref.~\cite{SanataniShiraishi}, respectively.
These examples also illustrate the locality of the conserved quantities.
More generally, since the degree of $f^{(n)}$ is bounded by $n$,
$Q^{(n)}$ is local for $1\leq n\leq L/2$.

\section{\label{sec:generalize}Generalizations}
We next generalize the construction to more general interactions and hopping terms.
We first consider extending interactions beyond the on-site form.
We then turn to cubic interactions, for which the same strategy applies, but requires different trigonometric identities.
Finally, we show that the construction can be extended to asymmetric hopping.

\subsection{\label{subsec:ex-int}Extended interactions}
As an example of an extended interaction, we consider the Hamiltonian
\begin{align}
    &H = T + V,\notag\\
    &T = \sum_{j=1}^{L}\qty(b^\dagger_jb_{j+1} + b^\dagger_{j+1}b_{j}), \quad 
        V= c\sum_{j=1}^{L}b^\dagger_jb^\dagger_{j+1}b^\dagger_{j+2}b^\dagger_{j+3},
\end{align}
which is written in the Fourier space as
\begin{align}
    T &= \sum_{p}2\cos(p)n_p,\\
    V &= \frac{c}{L}\sum_{p_1,p_2,p_3,p_4}\delta_{p_1+p_2+p_3+p_4,0}\,e^{-i(p_2+2p_3+3p_4)}\notag\\
    &\qquad\qquad \qquad \times b^\dagger_{p_1}b^\dagger_{p_2}b^\dagger_{p_3}b^\dagger_{p_4}.
\end{align}
The conserved quantities are then obtained as
\begin{align}
    Q^{(n)} &= T^{(n)} + V^{(n)},\quad 1\leq n\leq L/2,\notag\\
    T^{(n)} &= \sum_{p}\epsilon^{(n)}(p)n_p,\label{eq:qn4-ex-int}\\
    V^{(n)} &= \frac{c}{L}\sum_{p_1,p_2,p_3,p_4}e^{-i(p_2+2p_3+3p_4)}g^{(n)}(p_1,p_2,p_3,p_4)\notag\\
    &\qquad\qquad\times\delta_{p_1+p_2+p_3+p_4,0}\,b^\dagger_{p_1}b^\dagger_{p_2}b^\dagger_{p_3}b^\dagger_{p_4},\notag
\end{align}
where $\epsilon^{(n)}$ and $g^{(n)}$ are given in Eqs.~\eqref{eq:en4} and \eqref{eq:gn4}.
The construction can be extended in the same way to other 
quartic interactions of creation operators.

The extension beyond on-site interactions is also essential for fermionic systems, since the on-site quartic interaction vanishes identically for fermions.
The corresponding Hamiltonian and conserved quantities are obtained by replacing bosonic operators $b^\dagger_j$ and $b_j$ with fermionic operators $c^\dagger_j$ and $c_j$, respectively.

\subsection{\label{subsec:3body}Cubic interactions}
We next consider a model with cubic interactions.
The Hamiltonian is given by
\begin{align}
    &H = T+V,\notag\\
    &T = \sum_{j=1}^{L}\qty(b^\dagger_jb_{j+1} + b^\dagger_{j+1}b_{j}), \quad 
        V= c\sum_{j=1}^{L}\qty(b^\dagger_j)^3.
\end{align}
The conserved quantities are constructed in the same manner as for quartic interactions.
In the Fourier space, the Hamiltonian is written as
\begin{align}
    T &= \sum_{p}2\cos(p)n_p,\\
    V &= \frac{c}{\sqrt{L}}\sum_{p_1,p_2,p_3}\delta_{p_1+p_2+p_3,0}\,b^\dagger_{p_1}b^\dagger_{p_2}b^\dagger_{p_3}.
\end{align}
The conserved quantities are given by
\begin{align}
    Q^{(n)} &= T^{(n)} + V^{(n)},\quad 1\leq n\leq L/2,\notag\\
    T^{(n)} &= \sum_{p}\epsilon^{(n)}(p)n_p,\label{eq:Qn3}\\
    V^{(n)} &= \frac{c}{\sqrt{L}}\sum_{p_1,p_2,p_3}g^{(n)}(p_1,p_2,p_3)\notag\\
&\qquad\qquad\times\delta_{p_1+p_2+p_3,0}\,b^\dagger_{p_1}b^\dagger_{p_2}b^\dagger_{p_3},\notag
\end{align}
where
\begin{equation}
    g^{(n)}(p_1,p_2,p_3) = \frac{\sum_{l=1}^{3}\epsilon^{(n)}(p_l)}{\sum_{l=1}^{3}2\cos(p_l)}
\end{equation}
and $\epsilon^{(n)}$ is constructed such that $g^{(n)}$ is a trigonometric polynomial.
The construction of $\epsilon^{(n)}$ and the proof of this property are given in Appendix~\ref{app:cubic}.

\subsection{\label{subsec:asym}Asymmetric hopping}
We also consider asymmetric hopping with a cubic interaction:
\begin{equation}
    T = \sum_{j=1}^{L}\qty(b^\dagger_jb_{j+1} + \nu b^\dagger_{j+1}b_j),\quad V = c\sum_{j=1}^{L}\qty(b^\dagger_j)^3,\label{eq:ham-asym}
\end{equation}
where $\nu$ is an arbitrary complex number.
The conserved quantities are again constructed in the form of Eq.~\eqref{eq:Qn3}, with different functions $\epsilon^{(n)}$ and $g^{(n)}$.
Here, $g^{(n)}$ is defined as
\begin{equation}
    g^{(n)}(p_1,p_2,p_3) = \frac{\sum_{l=1}^3\epsilon^{(n)}(p_l)}{\sum_{l=1}^3\qty(e^{ip_l}+\nu e^{-ip_l})}\label{eq:gn3}
\end{equation}
and $\epsilon^{(n)}$ is constructed such that $g^{(n)}$ is a trigonometric polynomial.
For illustration, we explicitly give $\epsilon^{(n)}$ for the first few values of $n$:
\begin{align}
    \epsilon^{(2)}(p) = e^{2ip}-\nu^2e^{-2ip} + 2(\nu^3+1)e^{-ip},
\end{align}
\begin{align}
    \epsilon^{(3)}(p) &= e^{3ip} + \nu^3e^{-3ip} - 3\nu(\nu^3+1)e^{-2ip} \notag\\&+ 6\nu^2(\nu^3+1)e^{-ip} - (\nu^3+1),
\end{align}
\begin{align}
    \epsilon^{(4)}(p) &= e^{4ip}-\nu^4e^{-4ip} + 4\nu^2(\nu^3+1)e^{-3ip} \notag\\&- 2(5\nu^3+1)(\nu^3+1)e^{-2ip}\notag\\& + 4\nu(5\nu^3+2)(\nu^3+1)e^{-ip}-4\nu^2(\nu^3+1).
\end{align}
The construction of $\epsilon^{(n)}$ is given in Appendix~\ref{app:cubic}.
We note that, for $\nu=1$, these expressions reduce to those obtained for symmetric hopping in the previous subsection.
For $\nu = -1$, we obtain the simple form:
\begin{equation}
    \epsilon^{(n)}(p) = 2i\sin(np),\quad g^{(n)}(p_1,p_2,p_3) = \frac{\sum_{l=1}^{3}\sin(np_l)}{\sum_{l=1}^{3}\sin(p_l)}.
\end{equation}

\subsection{\label{subsec:chem}On-site potential}
Finally, we consider models with an on-site potential.
For cubic interactions,
$T$ in the Hamiltonian \eqref{eq:ham-asym} is modified as
\begin{equation}
    T = \sum_{j=1}^{L}\qty(b^\dagger_jb_{j+1} + \nu b^\dagger_{j+1}b_j + \mu b^\dagger_jb_j).\label{eq:chem3}
\end{equation}
In this case, the construction remains valid in the presence of an on-site potential.
This is consistent with the result of Ref.~\cite{SanataniShiraishi}, where the case $\nu=1$ is considered.

On the other hand, for quartic interactions, the situation is different.
We consider the Hamiltonian
\begin{align}
    H &= T+V,\notag\\
    T &= \sum_{j=1}^{L}\qty(b^\dagger_j b_{j+1} + b^\dagger_{j+1}b_j + \mu b^\dagger_jb_j),\quad V = c\sum_{j=1}^{L}\qty(b^\dagger_j)^4.
\end{align}
In this case, while we can construct a conserved quantity $Q^{(2)}$ with
\begin{equation}
    \epsilon^{(2)}(p) = e^{2ip} - e^{-2ip} + 4\mu(e^{ip} -e^{-ip}),
\end{equation}
the construction generally fails for $n\geq 3$.
This model corresponds to Type $\mathrm{N}^+$ in Ref.~\cite{SanataniShiraishi},
and our $Q^{(2)}$ coincides with their $Q_3$.
The obstruction to extending this construction is discussed in Appendix~\ref{app:quartic}.

\section{\label{sec:conclusion}Conclusion and outlook}
In this work, we have provided a new representation of the local conserved quantities previously constructed by Sanatani and Shiraishi in Ref~\cite{SanataniShiraishi} for a bosonic chain with quartic interactions of creation operators.
Using Fourier transformation and trigonometric identities, we obtained a systematic expression for the conserved quantities
and showed that these quantities form a mutually commuting family.
We further extended the construction to other forms of interactions and hopping.

It remains an open question whether the conserved quantities investigated here are related to an underlying Yang--Baxter structure.
More generally, it would be interesting to formulate the conserved quantities within a matrix-product-operator (MPO) framework
and investigate their commutation with the Hamiltonian using local conditions on the MPO tensors \cite{RubioMolnarSchuchVerstraete}.
MPO representations of conserved quantities are well established in Yang--Baxter integrable systems, where they are often in connection with transfer matrices \cite{FukaiYamada,YamadaFukai,Katsura,FendleyGehrmannVernierVerstraete,Yashin,Prosen,ProsenIlievski}.
However, an MPO representation by itself does not require Yang--Baxter integrability.
This suggests that an MPO formulation may provide a perspective complementary to the Yang--Baxter approach.

It would also be interesting to determine the most general class of interactions and hopping terms for which the present construction applies.
The construction presented in this paper relies crucially on the mutual commutativity of the momentum occupation operators $n_p$.
One direction is to extend to more general quadratic bosonic operators involving anomalous terms such as $b_pb_q$ and $b^\dagger_pb^\dagger_q$.
Although these operators no longer form a mutually commuting set,
general quadratic forms of bosonic operators are closed under commutation and form a Lie algebra \cite{SuMartinClerk,Auerbach}.
This algebraic structure may provide a natural starting point for extending the present approach to a broader class of bosonic models.
In particular, this extension can be applied to models with quadratic terms in creation operators and on-site potentials, including those belonging to Type $C^-$ in Ref.~\cite{SanataniShiraishi}.

Furthermore, the structure of the eigenstates remains largely unexplored.
In particular, constructing normalizable right eigenstates within the Fock space is difficult, since the interaction increases the particle number.
We therefore focus here on the construction of left eigenstates, for which several classes of eigenstates can be identified.
For one-, two-, and three-particle states, the quartic interaction term vanishes identically, so that the eigenstates reduce to those of the hopping Hamiltonian.
For higher particle numbers, one can construct left eigenstates by choosing momentum configurations for which the interaction term vanishes,
but a systematic characterization of such configurations is not known.
Moreover, since particle number is not conserved,
more general left eigenstates, including superpositions of states with different particle numbers, may also exist.
Related constructions of such states have been discussed in Ref.~\cite{AlcarazDrozHenkelRittenberg,NakagawaKawakamiUeda}.
A systematic understanding of the eigenstate structure is therefore an interesting direction for future work.

\begin{acknowledgments}
We would like to thank Naohisa Hirata and Naoto Shiraishi for fruitful discussions.
T.I. was supported JST SPRING, Grant Number JPMJSP2108, and Forefront Physics and Mathematics Program to Drive Transformation (FoPM), a World-leading
Innovative Graduate Study (WINGS) Program, the University of Tokyo. 
H.K. was supported by JSPS KAKENHI Grants No. JP23K25783 and No. 23K25790. 
\end{acknowledgments}

\section*{AI-use disclosure}
The authors used Gemini 1.5 (Google) for inspiring the core concept of this paper and ChatGPT-5.6 Luna (OpenAI) for assistance with the writing and editing of the manuscript. All scientific reasoning, data analysis, physical interpretations, and final text were directed, verified, and approved by the authors.

\appendix

\section{\label{app:quartic}Divisibility of trigonometric polynomials for quartic interactions}
In this appendix, we prove trigonometric identities in Eq.~\eqref{eq:trig-form-1} and show that $g^{(n)}$ in Eq.~\eqref{eq:gn4} is a trigonometric polynomial.

\subsection{\label{subsec:Newton-4}The Newton--Girard formulas}
To this end, we introduce the following notation.
We set $z_l = e^{ip_l}$ for $l=1,\ldots,4$
and introduce the elementary symmetric polynomials
\begin{align}
    &\mathsf{e}_1 = \sum_{l=1}^{4}z_l,\quad \mathsf{e}_2 = \!\sum_{1 \le l<m \le 4}\!z_lz_m,\quad \mathsf{e}_3 = \!\sum_{1 \le k<l<m \le 4}\!z_kz_lz_m,\notag\\
    &\mathsf{e}_4 = z_1z_2z_3z_4,\quad \mathsf{e}_k = 0,\quad(k>4),
\end{align}
and the power sums
\begin{equation}
    \mathsf{p}_n = \sum_{l=1}^{4}z_l^n.
\end{equation}
In terms of these polynomials, the sum of $\epsilon^{(n)}$ in Eq.~\eqref{eq:en4} is expressed as
\begin{equation}
    \sum_{l=1}^{4}\epsilon^{(n)}(p_l) = \mathsf{p}_n - (-1)^n\bar{\mathsf{p}}_n,
\end{equation}
where the bar denotes complex conjugation.
In particular, 
\begin{equation}
    \sum_{l=1}^{4}2\cos(p_l) = \mathsf{p}_1 + \bar{\mathsf{p}}_1.
\end{equation}
When the momentum conservation $p_1 + p_2 + p_3 + p_4 = 0$ (mod $2\pi$) is satisfied,
the elementary symmetric polynomials obey
\begin{equation}
    \bar{\mathsf{e}}_1 = \mathsf{e}_3,\quad \bar{\mathsf{e}}_2 = \mathsf{e}_2,\quad \mathsf{e}_4 = 1.\label{eq:moment-conserve}
\end{equation}
The power sums and elementary symmetric polynomials obey the Newton--Girard formulas \cite{Mcdonald}:
\begin{equation}
    \mathsf{p}_n = \sum_{k=1}^{n-1}(-1)^{k-1}\mathsf{e}_k\mathsf{p}_{n-k} - (-1)^nn\mathsf{e}_n.\label{eq:Newton}
\end{equation}
By applying Eq.~\eqref{eq:Newton} recursively, we obtain
\begin{equation}
    \mathsf{p}_n = \mathsf{q}_n(\mathsf{e}_1,\mathsf{e}_2,\mathsf{e}_3),
\end{equation}
where $\mathsf{q}_n(x_1,x_2,x_3)$ is a polynomial.
Here, we have used the momentum-conservation condition \eqref{eq:moment-conserve}, which gives $\mathsf{e}_4=1$ and allows us to omit $\mathsf{e}_4$ as an independent variable.
For the first few orders, we have
\begin{align}
    \mathsf{q}_2(\mathsf{e}_1,\mathsf{e}_2,\mathsf{e}_3) &= \mathsf{e}_1^2 - 2\mathsf{e}_2,\\
    \mathsf{q}_3(\mathsf{e}_1,\mathsf{e}_2,\mathsf{e}_3) &= \mathsf{e}_1^3 - 3\mathsf{e}_1\mathsf{e}_2 + 3\mathsf{e}_3,\\
    \mathsf{q}_4(\mathsf{e}_1,\mathsf{e}_2,\mathsf{e}_3) &= \mathsf{e}_1^4 - 4\mathsf{e}_1^2\mathsf{e}_2 + 4\mathsf{e}_1\mathsf{e}_3 + 2\mathsf{e}_2^2 - 4.
\end{align}

\subsection{\label{subsec:proof4}Proof of trigonometric identities}
We now turn to the proof of Eq.~\eqref{eq:trig-form-1}.
With the momentum-conservation condition \eqref{eq:moment-conserve}, the complex conjugate of $\mathsf{p}_n$ is obtained by exchanging $\mathsf{e}_3$ and $\mathsf{e}_1$:
\begin{equation}
    \bar{\mathsf{p}}_n(\mathsf{e}_1,\mathsf{e}_2,\mathsf{e}_3) = \mathsf{q}_n(\mathsf{e}_3,\mathsf{e}_2,\mathsf{e}_1).
\end{equation}
Thus, it suffices to show that $\mathsf{q}_n(\mathsf{e}_1,\mathsf{e}_2,\mathsf{e}_3) - (-1)^n\mathsf{q}_n(\mathsf{e}_3,\mathsf{e}_2,\mathsf{e}_1)$ is divisible by $\mathsf{p}_1 +\bar{\mathsf{p}}_1 = \mathsf{e}_1+\mathsf{e}_3$.
Equivalently, by the factor theorem,
it suffices to show that
\begin{equation}
    \mathsf{q}_n(x,y,-x) - (-1)^{n}\mathsf{q}_n(-x,y,x) 
    = 0\label{eq:condition-4}
\end{equation}
as a polynomial in $x,y$.

We first establish a parity property of the polynomial $\mathsf{q}_n$.
In every monomial appearing in $\mathsf{q}_n(\mathsf{e}_1,\mathsf{e}_2,\mathsf{e}_3)$,
the total degree in $\mathsf{e}_1$ and $\mathsf{e}_3$ has the same parity as $n$.
This property follows by mathematical induction from the Newton--Girard formulas.
For $n=1,2,3,4$, the property can be verified directly.
Now, suppose that the property holds for all $\mathsf{q}_m$ with $m<n$, and consider $n>4$.
Since $\mathsf{e}_k = 0$ for $k>4$, every nonzero term $\mathsf{e}_k\mathsf{p}_{n-k}$ in Eq.~\eqref{eq:Newton} has $1\leq k\leq 4$.
For $k=1,2,3,4$, the degrees of $\mathsf{e}_k$ in $\mathsf{e}_1$ and $\mathsf{e}_3$ are $1,0,1,0$, respectively, and hence have the same parity as $k$.
Therefore, every monomial in $\mathsf{e}_k\mathsf{p}_{n-k}$ has total degree in $\mathsf{e}_1$ and $\mathsf{e}_3$ congruent to
\begin{equation}
    k + (n-k) = n\quad (\mathrm{mod}~2).
\end{equation}
Hence the property holds for $\mathsf{q}_n$, and thus for all $n$.

We can now prove Eq.~\eqref{eq:condition-4}.
Consider a monomial $(\mathsf{e}_1)^a(\mathsf{e}_2)^b(\mathsf{e}_3)^c$ in $\mathsf{q}_n(\mathsf{e}_1,\mathsf{e}_2,\mathsf{e}_3)$.
By the parity property established above, $a+c$ has the same parity as $n$.
Therefore,
\begin{equation}
    x^a(-x)^c = (-1)^{a+c}(-x)^ax^c = (-1)^n(-x)^ax^c,
\end{equation}
and consequently Eq.~\eqref{eq:condition-4} follows.
By the factor theorem, it follows that
\begin{equation}
    \sum_{l=1}^{4}\epsilon^{(n)}(p_l) = (\mathsf{e}_1+\mathsf{e}_3)\,\mathsf{f}_n(\mathsf{e}_1,\mathsf{e}_2,\mathsf{e}_3),\label{eq:fn4}
\end{equation}
where $\mathsf{f}_n$ is a polynomial in $\mathsf{e}_1,\mathsf{e}_2$, and $\mathsf{e}_3$,
which proves Eq.~\eqref{eq:trig-form-1}.
We note that $\mathsf{f}_n$ gives the trigonometric polynomial $f^{(n)}$ in Eq.~\eqref{eq:trig-form-1} as a function of $p_1,\ldots, p_4$:
\begin{equation}
    \mathsf{f}_n(\mathsf{e}_1,\mathsf{e}_2,\mathsf{e}_3) = f^{(n)}(p_1,p_2,p_3,p_4).
\end{equation}

To characterize the degree of $f^{(n)}$,
we assign weights to the elementary symmetric polynomials according to their degrees as Laurent polynomials in $z_l$ under the momentum-conservation condition \eqref{eq:moment-conserve}:
\begin{equation}
    \deg\mathsf{e}_1 = \deg\mathsf{e}_3 = 1,\quad \deg\mathsf{e}_2 = 2,\quad \deg\mathsf{e}_4 = 0.
\end{equation}
With these assignments,
the left-hand side of Eq.~\eqref{eq:fn4} has degree $n$, while $\mathsf{e}_1+\mathsf{e}_3$ has degree $1$.
It follows that $f^{(n)}$ has degree at most $n-1$.
Since $g^{(n)} = f^{(n)}$ follows from Eqs.~\eqref{eq:gn4} and \eqref{eq:trig-form-1}, $g^{(n)}$ is a trigonometric polynomial with degree at most $n-1$.

All the functions $f^{(n)}$ can be encoded in a single generating function 
\begin{equation}
    G(u; p_1, p_2, p_3, p_4) = \sum_{n=1}^{\infty}u^{n-1}f^{(n)}(p_1, p_2, p_3, p_4),
\end{equation}
which admits an explicit expression in terms of $\mathsf{e}_1$, $\mathsf{e}_2$, and $\mathsf{e}_3$ as 
\begin{widetext}
\begin{align}
G(u;p_1,p_2,p_3,p_4)= \frac{1 + (3-\mathsf{e}_2)u^2 - 2(\mathsf{e}_1-\mathsf{e}_3)u^3 + (\mathsf{e}_2-3)u^4 - u^6}{(1-u\mathsf{e}_1+u^2\mathsf{e}_2-u^3\mathsf{e}_3+u^4)(1+u\mathsf{e}_3+u^2\mathsf{e}_2+u^3\mathsf{e}_1+u^4)}.
  \end{align}
\end{widetext}

\subsection{\label{subsec:chem-4}Obstruction from an on-site potential}
We finally consider the effect of an on-site potential on the construction for quartic interactions.
In this case, the denominator of the right-hand side of Eq.~\eqref{eq:gn4} is modified to
\begin{equation}
    \sum_{l=1}^{4}\qty[2\cos(p_l) + \mu] = \mathsf{e}_1+\mathsf{e}_3 + 4\mu.
\end{equation}
Thus, for the numerator
\begin{equation}
\mathsf{r}_n(\mathsf{e}_1,\mathsf{e}_2,\mathsf{e}_3) = \sum_{l=1}^{4}\epsilon^{(n)}(p_l),
\end{equation}
the factor theorem requires
\begin{equation}
    \mathsf{r}_n(x-4\mu,\mathsf{e}_2,-x)= 0.\label{eq:factor-4-chem}
\end{equation}

For $\mu = 0$, this condition is satisfied by the parity property established above.
Indeed, every monomial in $\mathsf{q}_n$ has the same parity in the total degree of $\mathsf{e}_1$ and $\mathsf{e}_3$ as $n$,
which leads to the cancellation in Eq.~\eqref{eq:condition-4}.
For $\mu\neq 0$, however, this parity argument no longer applies.
Under the substitution $\mathsf{e}_1 = x-4\mu$ and $\mathsf{e}_3 = -x$,
the terms generated by the shift of $\mathsf{e}_1$ generally prevent the resulting polynomial from vanishing identically.
For example, starting from $n=3$, terms involving $\mu\mathsf{e}_2$ remain and cannot be canceled by the available lower-order terms.
Thus, Eq.~\eqref{eq:factor-4-chem} cannot in general be satisfied for $n\geq 3$.
An exception is the case of $n=2$.
In this case, we can choose
\begin{align}
    \mathsf{r}_2(\mathsf{e}_1,\mathsf{e}_2,\mathsf{e}_3) &= \mathsf{q}_2(\mathsf{e}_1,\mathsf{e}_2,\mathsf{e}_3)-\mathsf{q}_2(\mathsf{e}_3,\mathsf{e}_2,\mathsf{e}_1) \\
    &+ 4\mu\qty(\mathsf{q}_1(\mathsf{e}_1,\mathsf{e}_2,\mathsf{e}_3)-\mathsf{q}_1(\mathsf{e}_3,\mathsf{e}_2,\mathsf{e}_1)),
\end{align}
for which
\begin{equation}
    \mathsf{r}_2(x-4\mu,\mathsf{e}_2,-x) = 0.
\end{equation}
The cancellation is possible because the $\mathsf{e}_2$ terms in $\mathsf{p}_2-\bar{\mathsf{p}}_2$ cancel identically.
This gives the choice of $\epsilon^{(2)}(p)$ in the main text and the corresponding conserved quantity $Q^{(2)}$.

\section{\label{app:cubic}Construction of $\epsilon^{(n)}$ for cubic interactions}
In this appendix, we construct $\epsilon^{(n)}$ for the system with cubic interactions and asymmetric hopping,
described by the Hamiltonian \eqref{eq:ham-asym}.
As in App.~\ref{app:quartic}, we set $z_l = e^{ip_l}$ for $l=1,\ldots,3$ and introduce the elementary symmetric polynomials
\begin{align}
    &\mathsf{e}_1 = \sum_{l=1}^{3}z_l,\quad \mathsf{e}_2 = \!\sum_{1\le l<m \le 3}\!z_lz_m,\quad \mathsf{e}_3 = z_1z_2z_3,\notag\\
    &\mathsf{e}_k = 0, \quad(k>3),
\end{align}
and the power sums
\begin{equation}
    \mathsf{p}_n = \sum_{l=1}^{3}z_l^n.
\end{equation}
When the momentum conservation $p_1 + p_2 + p_3 = 0$ (mod $2\pi$) is satisfied, we have
\begin{equation}
    \bar{\mathsf{e}}_1 = \mathsf{e}_2,\quad\mathsf{e}_3=1.\label{eq:moment-cons-3}
\end{equation}

We now construct $\epsilon^{(n)}$ such that the right-hand side of Eq.~\eqref{eq:gn3} is a trigonometric polynomial.
The denominator of Eq.~\eqref{eq:gn3} is written as
\begin{equation}
    \sum_{l=1}^{3}\qty(e^{ip_l} + \nu e^{-ip_l}) = \mathsf{e}_1 + \nu \mathsf{e}_2.
\end{equation}
Using the Newton--Girard formulas in Eq.~\eqref{eq:Newton}, the power sums $\mathsf{p}_n$ can be expressed as
\begin{equation}
    \mathsf{p}_n = \mathsf{q}_n(\mathsf{e}_1,\mathsf{e}_2),\quad \bar{\mathsf{p}}_n = \mathsf{q}_n(\mathsf{e}_2,\mathsf{e}_1),
\end{equation}
where $\mathsf{q}_n(x_1,x_2)$ is a polynomial.
Here, we have used the momentum-conservation condition \eqref{eq:moment-cons-3}.
For example, we have
\begin{align}
    \mathsf{q}_2(\mathsf{e}_1,\mathsf{e}_2) &= \mathsf{e}_1^2 - 2\mathsf{e}_2,\\
    \mathsf{q}_3(\mathsf{e}_1,\mathsf{e}_2) &= \mathsf{e}_1^3 - 3\mathsf{e}_1\mathsf{e}_2 + 3,\\
    \mathsf{q}_4(\mathsf{e}_1,\mathsf{e}_2) &= \mathsf{e}_1^4 - 4\mathsf{e}_1^2\mathsf{e}_2 + 4\mathsf{e}_1 + 2\mathsf{e}_2^2.
\end{align}
We seek a polynomial $\mathsf{r}_n(\mathsf{e}_1,\mathsf{e}_2)$ of the form
\begin{equation}
    \mathsf{r}_n(\mathsf{e}_1,\mathsf{e}_2) = \sum_{l=1}^{3}\epsilon^{(n)}(p_l)\label{eq:def-en-3}
\end{equation}
that is a linear combination of $\mathsf{p}_1,\ldots,\mathsf{p}_n$ and $\bar{\mathsf{p}}_1,\ldots,\bar{\mathsf{p}}_n$, and is divisible by $\mathsf{e}_1 + \nu\mathsf{e}_2$.
By the factor theorem, this is equivalent to requiring
\begin{equation}
    \mathsf{r}_n(\nu x,-x)=0.
\end{equation}

We construct $\mathsf{r}_n$ successively, starting from
\begin{equation}
    \mathsf{r}_n^{(0)}(\mathsf{e}_1,\mathsf{e}_2) = \mathsf{q}_n(\mathsf{e}_1,\mathsf{e}_2) - (-1)^n\nu^n\mathsf{q}_n(\mathsf{e}_2,\mathsf{e}_1).
\end{equation}
The terms of degree $n$ cancel under $\mathsf{e}_1 = \nu x$ and $\mathsf{e}_2 = -x$,
so that $\mathsf{r}_n^{(0)}(\nu x,-x)$ has degree less than $n$ as a polynomial in $x$.
We can therefore add an appropriate multiple of $\mathsf{q}_{n-1}(\mathsf{e}_2,\mathsf{e}_1)$ to cancel the highest remaining degree.
Repeating this procedure, we obtain a polynomial
\begin{equation}
    \mathsf{r}_n(\mathsf{e}_1,\mathsf{e}_2) = \mathsf{r}_n^{(0)}(\mathsf{e}_1,\mathsf{e}_2) + \sum_{m<n}a_m\mathsf{q}_m(\mathsf{e}_2,\mathsf{e}_1)
\end{equation}
such that
\begin{equation}
    \mathsf{r}_n(\nu x, -x) = 0.
\end{equation}
The coefficients $a_m$ are determined successively by canceling the remaining powers of $x$ from the highest degree to the lowest.
We note that this construction is not unique, since linear combinations of $\mathsf{r}_m$ ($m<n$) can be added to $\mathsf{r}_n$.
This ambiguity corresponds to the freedom to add lower-order conserved quantities to $Q^{(n)}$.

The same successive construction also applies when the denominator is shifted by a constant.
In particular, for a denominator of the form $\mathsf{e}_1 + \nu\mathsf{e}_2 + 3\mu$, one imposes $\mathsf{e}_1 = \nu x - 3\mu$ and $\mathsf{e}_2 = -x$,
and the coefficients can again be chosen successively to cancel the resulting polynomial in $x$.
This denominator corresponds to the Hamiltonian with $T$ given in Eq.~\eqref{eq:chem3} and the cubic interaction.

\bibliography{reference}

\end{document}